\documentclass[aps,prl, amsmath,showkeys,floatfix,amssymb, preprintnumbers, nofootinbib, superscriptaddress,reprint]{revtex4-1}
\usepackage{graphicx,color}
\usepackage{amsmath,dsfont}
\usepackage{amssymb}
\usepackage{booktabs}
\usepackage{mathrsfs}
\usepackage{cases}
\usepackage[colorlinks=true,linkcolor=blue,plainpages=false]{hyperref}
\usepackage[capitalise]{cleveref}
\usepackage{chngcntr}
\usepackage{floatrow}
\def\simle{\mathrel{\rlap{\raise 0.511ex \hbox{$<$}}{\lower 0.511ex \hbox{$\sim$}}}}
\def\simge{\mathrel{ \rlap{\raise 0.511ex \hbox{$>$}}{\lower 0.511ex \hbox{$\sim$}}}}

\newcommand \beq{\begin{flalign}}
\newcommand \eeq{\end{flalign}} 
\def\simle{\mathrel{\rlap{\raise 0.511ex \hbox{$<$}}{\lower 0.511ex 
\hbox{$\sim$}}}}
\def\simge{\mathrel{ \rlap{\raise 0.511ex 
\hbox{$>$}}{\lower 0.511ex \hbox{$\sim$}}}}

\newcommand\lb{\left(}
\newcommand\rb{\right)}
\newcommand{\lsb}{\left[ }
\newcommand{\rsb}{\right]}

\newcommand*\diff{\mathop{}\!\mathrm{d}}
\newcommand*\Diff[1]{\mathop{}\!\mathrm{d^#1}}
\newcommand{\Tr}{{\rm Tr}}

\begin{document}

\title{Confinement from Correlated Instanton-Dyon Ensemble in SU(2) Yang-Mills Theory}
\medskip
 
 \author{Miguel Angel Lopez-Ruiz}
 \email{malopezr@iu.edu}
\affiliation{ Physics Department and Center for Exploration of Energy and Matter,
Indiana University, 2401 N Milo B. Sampson Lane, Bloomington, IN 47408, USA.}

\author{Yin Jiang}
\email{jiang\_y@buaa.edu.cn}
\affiliation{Physics Department, Beihang University, 37 Xueyuan Rd, Beijing 100191, China.}

\author{Jinfeng Liao}
\email{liaoji@indiana.edu}
\affiliation{ Physics Department and Center for Exploration of Energy and Matter,
Indiana University, 2401 N Milo B. Sampson Lane, Bloomington, IN 47408, USA.}

\date{\today}

\begin{abstract}
	We study the  confinement phase transition in $SU(2)$ Yang-Mills theory, based on a statistical ensemble model of correlated instanton-dyons. We show for the first time that such a  model provides a quantitative description, in light of the  lattice data, for the temperature dependence of the order parameter. We characterize the short-range interaction  which plays a  crucial role for the properties of such ensemble.  The chromo-magnetic charge density as well as the spatial correlations is found to be consistent with known lattice and phenomenological information. 

%
\end{abstract}

\maketitle


{\em Introduction.---} Sixty-five years after the advent of Yang-Mills theory~\cite{Yang:1954ek} and more than forty-five years after the discovery of Quantum Chromodynamics (QCD)~\cite{Gross:1973id,Politzer:1973fx}, an understanding of the mechanism for confinement phenomenon in such theories remains a significant challenge~\cite{Greensite:2011zz,Alkofer:2006fu}. First-principle lattice simulations have proven that confinement is indeed a consequence of the underlying gluon fields in the strongly coupled regime and provided ample detailed information about the transition between confined and deconfined phases~\cite{Greensite:2011zz,Gattringer:2010zz,Kondo:2014sta}. Heavy ion collision experiments at the  Relativistic Heavy Ion Collider (RHIC) and the Large Hadron Collider (LHC)   have also allowed phenomenologically extracting  many properties of hot matter in the vicinity of the confinement transition~\cite{Shuryak:2004cy,Gyulassy:2004zy,Muller:2012zq,Shuryak:2014zxa}.  Nevertheless, we do not have a precise picture of how confinement occurs and  what are the relevant degrees of freedom driving this phenomenon.

Recently, a promising approach has emerged; based on a new class of gluon topological configurations known as the instanton-dyons~\cite{Kraan:1998pm,Lee:1998bb,Diakonov:2004jn,Diakonov:2007nv,Diakonov:2009jq,Shuryak:2013tka,Faccioli:2013ja,Larsen:2015vaa,Larsen:2015tso,Liu:2015ufa,Liu:2015jsa,Ramamurti:2018evz,Lopez-Ruiz:2016bjl}. This paper aims to provide, for the first time, a quantitative description of the confinement transition in $SU(2)$ Yang-Mills theory based on this approach.  In the following, we will first formulate the confinement problem and introduce the instanton-dyon ensemble model in an accessible way. We will then present detailed results to be compared with lattice data as well as discuss the phenomenological implications. 

{\em Holonomy Potential.---} Let us start by formulating the confinement problem in pure Yang-Mills theory in terms of the holonomy potential.  In the imaginary time formalism for finite temperature field theory, one can define the Polyakov loop for a  gauge configuration $A_\mu$ as:  
\begin{eqnarray}
\mathcal{L}[A_\mu]=\hat{\mathcal{P}}\ {\rm exp}\left(i\int_0^{1/T} \diff x_4 \ A_4(\vec{x},x_4) \right),
\end{eqnarray}
where $T$ is temperature and $\hat{\mathcal{P}}$ is the usual path ordering. As is well known, the  gauge invariant expectation value $L\equiv \langle \frac{1}{N_c}\ \Tr \ \mathcal{L} \rangle$, often also simply referred to as the Polyakov loop, is a well-defined order parameter for confinement transition in pure Yang-Mills theories~\cite{Greensite:2011zz,Gattringer:2010zz}. The value of $L$ provides a measure of the ``penalty'': $L=0$ implies infinite free energy cost while $L=1$ implies zero cost  for creating a free color charge (in the fundamental representation). Therefore, in the confined phase below critical temperature $T_c$, one has $L =0 $, whereas at $T>T_c$ one has $L>0$  which approaches unity with increasing temperature. We note highly interesting analytical insights for the pertinent confinement dynamics in deformations of Yang-Mills theories on $R^3\times S^1$~\cite{Sulejmanpasic:2016uwq,Cherman:2016hcd,Poppitz:2013zqa,Poppitz:2012sw}. 

One can  classify all the gauge configurations according to the boundary values of the Polyakov loop, $\mathcal{L}_\infty \equiv \mathcal{L}[A_\mu]{\big |}_{|\vec{x}|\to \infty }$~\cite{Gross:1980br,Weiss:1980rj}. We focus on the $SU(2)$ case. Up to a gauge transformation and owing to the traceless nature of gauge group generators, one can always characterize the boundary values with one parameter $h \in [0,1]$: $\mathcal{L}_\infty = \text{diag}(e^{-i\pi h}, e^{i\pi h})$. Correspondingly, for configurations with such boundary values, one has 
\begin{eqnarray}
L_\infty = \frac{1}{2}\ \Tr \ \mathcal{L}_\infty =  \cos \left( \pi h \right)   \, .
\end{eqnarray}
The above gauge invariant value is the holonomy, with $h$ being the holonomy parameter.   For later convenience we also introduce $\bar{h}\equiv 1-h$.  The confining holonomy corresponds to $L_\infty=0$ thus $h=1/2$, while the trivial holonomy corresponds to  $L_\infty=1$ hence $h=0$. 

One can then classify all gauge configurations according to their holonomy values, and rewrite the path integral formulation of the theory's partition function as: 
\begin{eqnarray}
\mathcal{Z} &=& \int [\mathcal{D} A_\mu] e^{-S_E}  \nonumber \\
&&  \to  \int \diff h \ \left\{\int [\mathcal{D} A^h_\mu] e^{-S_E} \right \} = \int \diff h\  e^{-  \mathcal{U}[h] V/T}
\end{eqnarray}
where $A^h_\mu$ are all gauge configurations with holonomy value $h$, $V$ is the system volume, $T$ is temperature and the potential $\mathcal{U}[h]$ or $\mathcal{U}[L_\infty]$ is the holonomy potential.

In the thermodynamic equilibrium at a given temperature $T$, the expectation value of the Polyakov loop $\langle  L_\infty \rangle $ must correspond to   the {\em minimum of the holonomy potential}. Therefore, by computing this holonomy potential and examining its minimum, one would be able to determine $\langle  L_\infty \rangle $ and its dependence on temperature. In this formulation of the confinement problem, the essential question is to reveal the shape of the potential $\mathcal{U}[L_\infty]$ and the holonomy value at its minimum. 

As a famous example, one could compute the (one-loop) perturbative contributions from gluons to the holonomy potential. This neat result, from Gross-Pisarski-Yaffe (GPY)~\cite{Gross:1980br,Weiss:1980rj} ,  is given by:
\begin{equation}
	\mathcal{U}^I_{GPY} = \frac{4\pi^2}{3}\  T^4 \ h^2 \bar{h}^2 \,\, .
\label{Eq:GPYI}
\end{equation} 
 It shall be obvious that the above perturbative potential has its minimum at $h=0$ or $h=1$, i.e. corresponding to trivial (non-confining) holonomy. That is, perturbative contributions can not lead to confinement. Contributions to holonomy potential that would be capable of changing its shape toward a minimum at confining holonomy (with $h=\bar{h}=1/2$), therefore, must come from non-perturbative sectors, as we shall discuss next.

{\em Correlated Instanton-Dyon Ensemble.---} It has been long suspected that an ensemble of gluonic topological configurations holds the key of confinement mechanism and their contributions to the holonomy potential should drive its minimum toward the non-trivial, confining value
~\cite{'tHooft:1981ht,tHooft:1982ylj,tHooft:1999cgx,Nambu:1974zg,Mandelstam:1974pi}. The hard question is what type of topological configurations would be the right degrees of freedom. They need to carry chromo-magnetic charges to be compatible with  the ``dual superconductor'' picture for confining vacuum, which appears to be supported by extensive lattice studies~\cite{Kondo:2014sta,Ripka:2003vv}. Their properties also need to be sensitive to holonomy in order to influence the behavior of the holonomy potential.  The instanton-dyons, which are constituents of the KvBLL calorons, appear to be the promising candidate satisfying both requirements. Let us briefly discuss these objects in the following.  

\begin{table}[!t]
	{\caption{Properties of the $SU(2)$ instanton-dyons.}\label{tab1}}
	\begin{ruledtabular}
	\begin{tabular}{*{5}{c}}
    & $M$ & $\bar{M}$ & $L$ & $\bar{L}$ \\
	\hline
	$\begin{array}{c}
	\text{Electric charge}
	\end{array}$ & 1 & 1 & -1 & -1 \\
	$\begin{array}{c}
	\text{Magnetic charge}
	\end{array}$ & 1 & -1 & -1 & 1 \\
	$\begin{array}{c}
	\text{h-charge}
	\end{array}$ & 1  & 1  & -1 & -1 \\
	\vspace*{2mm}
	Action & $h\frac{8\pi^2}{g^2}$ & $h\frac{8\pi^2}{g^2}$ & $\bar{h}\frac{8\pi^2}{g^2}$ & $\bar{h}\frac{8\pi^2}{g^2}$ \\
	Size & $(2\pi Th)^{-1}$ & $ (2\pi Th)^{-1}$ & $(2\pi T\bar{h})^{-1}$ & 
	$(2\pi T\bar{h})^{-1}$\\
	\end{tabular}
	\end{ruledtabular}
\end{table}

The KvBLL caloron, found relatively recently~\cite{Kraan:1998pm,Lee:1998bb}, is a new type of finite-temperature instanton solution  with non-trivial holonomy. See e.g.~\cite{Diakonov:2009jq} for reviews. The most remarkable feature is that  each such caloron of gauge group $SU( N_c)$ is made of $N_c$ constituents that are magnetically charged. These constituents are referred to as instanton-dyons. Specifically for the $SU(2)$ case, there are four types of instanton-dyons: the $L$- and $M$-dyon together making a KvBLL caloron while the $\bar{L}$- and $\bar{M}$-dyon make an anti-caloron.  The key properties of the instanton-dyons are summarized in \cref{tab1}. While a caloron  always has its action to be the familiar  $8\pi^2/g^2$ (with $g$ the gauge coupling) independent of holonomy, the division of this action between the two constituents as well as the size of these objects do sensitively depend on the holonomy parameter $h$. Even though a caloron is both electrically and magnetically neutral, its constituent dyons do carry non-zero charges.  These non-trivial features of instanton-dyons have generated hope that  confinement could be explained by their contributions. A number of analytic and numerical studies have shown results in strong support of such a scenario~\cite{Diakonov:2004jn,Diakonov:2007nv,Diakonov:2009jq,Shuryak:2013tka,Faccioli:2013ja,Larsen:2015vaa,Larsen:2015tso,Liu:2015ufa,Liu:2015jsa,Ramamurti:2018evz,Lopez-Ruiz:2016bjl}.

To investigate confinement, one needs to  compute the contributions of instanton-dyons to the holonomy potential. To do that, we build a statistical ensemble of these objects for any given holonomy value as follows:
\begin{widetext}
\begin{align}
\mathcal{Z}^{dyon}_h  = e^{-{\mathcal U}^{II}_{GPY}(h)\  V/T} & \sum_{\substack{N_M,N_L, \\ N_{\bar{L}}, N_{\bar{M}}}} \frac{1}{N_L!N_M!N_{\bar{L}}!N_{\bar{M}}!} \int \prod_{l=1}^{N_L} f_L T^3 \Diff{3} r_{L_l} \prod_{m=1}^{N_M} f_M T^3 \Diff{3} r_{M_m} \nonumber \\
  & \times \prod_{\bar{l}=1}^{N_{\bar{L}}}f_{\bar{L}} T^3 \Diff{3} r_{\bar{L}_{\bar{l}}} \prod_{\bar{m}=1}^{N_{\bar{M}}} f_{\bar{M}} T^3 \Diff{3} r_{\bar{M}_{\bar{m}}} \det(G_D)\det(G_{\bar{D}}) \,e^{-V_{D\bar{D}}},
\label{Eq:partition}
\end{align}
\end{widetext} 
The above sums over various configurations with $N_L$, $N_M$, $N_{\bar{L}}$, and $N_{\bar{M}}$ numbers of $L$-,$M$-, $\bar{L}$- and $\bar{M}$-dyons respectively. These objects are distributed over the spatial volume with their respective coordinates labeled by 
$r_{L_l}$, $r_{M_m}$, $r_{\bar{L}_{\bar{l}}}$, and $r_{\bar{M}_{\bar{m}}}$. The determinant terms $\det(G_D)$ and $\det(G_{\bar{D}})$ come from the   quantum weight for dyons and antidyons by computing one-loop quantum fluctuations around background fields of the calorons; the detailed form of which can be found in e.g. \cite{Diakonov:2004jn,Larsen:2015vaa,Lopez-Ruiz:2016bjl}. The $f_L$, $f_M$, $f_{\bar{L}}$, and $f_{\bar{M}}$ are the fugacity factors given by:
 \begin{flalign}
 \nonumber
 	f_M &= f_{\bar{M}} = S^2 \ e^{-h S}\ h^{\frac{8h}{3}-1} \,\, ,  \\ 
f_L &= f_{\bar{L}}= S^2\  e^{-\bar{h} S} \ \bar{h}^{\frac{8\bar{h}}{3}-1} \,\, .
	\label{Eq:fugacity}
 \end{flalign} 
Studies on instanton-dyon ensemble models of this sort were pioneered in~\cite{Shuryak:2013tka,Larsen:2015vaa}.  Various qualitative   features of such models were investigated in  ~\cite{Faccioli:2013ja,Larsen:2015tso,Ramamurti:2018evz,Lopez-Ruiz:2016bjl}.  

An important quantity in the partition function $\mathcal{Z}^{dyon}_h$  is the caloron action $S$, which is essentially the ``control parameter'' of the ensemble. While classically one simply has $S=8\pi^2/g^2$, quantum loop corrections render the coupling to run with temperature scale $T$. Here, next-to-leading order effects are considered by taking the two-loop correction to the gauge coupling \cite{Diakonov:2007nv}, thus defining the relation between the action and temperature via
\begin{equation}
	S(T)\approx\frac{22}{3}\log\lb\frac{T}{\Lambda}\rb + \frac{34}{11}\log\lsb 2\log\lb\frac{T}{\Lambda}\rb \rsb,
	\label{Eq:2LoopS}
\end{equation}
 where $\Lambda$ is the  non-perturbative scale. By varying $S$ from large to small values, the system changes from high to low temperature or equivalently from weak to strong coupling regime. In addition, we consistently include the two-loop correction to the perturbative potential \cref{Eq:GPYI}, which takes the simple and compact form \cite{Dumitru:2013xna}
 \begin{equation}
 	\mathcal{U}_{GPY}^{II}=\lb 1-\frac{5}{S} \rb \mathcal{U}_{GPY}^{I}.
 \end{equation}

A crucial ingredient for the properties of the ensemble is the interaction among the instanton-dyons within the ensemble. This is implemented via the  $V_{D\bar{D}}$ term in \cref{Eq:partition}. Such interaction has two features. At long distance, the interaction between a pair of constituents $i$ and $j$ at a spatial distance $r_{ij}$ should be a Coulomb force according to the objects' $e,m,h$ charges in \cref{tab1}:
\begin{equation}
	V_{long} =  \left(  e_i e_j + m_i m_j - 2 h_i h_j  \right) \, \frac{S}{2\pi T} \, \frac{e^{-M_D r_{ij}}}{r_{ij}}.
\end{equation}

  The screening effect in such a many-body ensemble of charges has been implemented through  a Debye mass parameter $M_D$ in the above. Note  that between an $L$-$M$ pair (and similarly $\bar{L}$-$\bar{M}$ pair), which together can make a full caloron, all interactions  cancel out by virtue of their BPS nature~\cite{Kraan:1998pm,Lee:1998bb,Diakonov:2004jn}. The correlations between these pairs are encoded in the determinant terms. In between an $L$-$\bar{M}$ or $\bar{L}$-$M$ pair, the Coulomb force is repulsive and prevents unphysical overlapping between them. For the other pair combinations (i.e. $L$-$L$, $\bar{L}$-$\bar{L}$, $L$-$\bar{L}$ as well as $M$-$M$, $\bar{M}$-$\bar{M}$, $M$-$\bar{M}$), a repulsive force at short distance needs to occur and stabilize the ensemble~\cite{Shuryak:2013tka}. We use the following short-range core-like interaction~\cite{Larsen:2015vaa,Larsen:2014yya}:
\begin{equation}
	V_{short}  =  \frac{c_h V_c}{1+e^{(2\pi T) r_{ij} c_h- \zeta_c}} ,  \,\,\,\, \text{for} \,\,\,\, r_{ij}  <   \frac{\zeta_c}{(2\pi T) c_h},
\end{equation}
where the coefficient $c_h=h$ for $M$-sector while $c_h=\bar{h}$ for $L$-sector, reflecting the different properties of the two sectors. $V_c$ is the strength parameter of this repulsive potential. $\zeta_c$ is the range parameter that separates the short and long-distance regions. The repulsive potential becomes important when the ensemble  becomes dense and it strongly influences the short-range correlations among constituents. The confining properties of such ensemble are sensitive to  the key parameters $V_c$ and $\zeta_c$~\cite{Lopez-Ruiz:2016bjl}.   

Our goal here is to investigate the viability of this effective description for confinement in light of  first-principle lattice calculation results and to characterize  the necessary parameters of such an ensemble in order to quantitatively describe the confinement transition in the $SU(2)$ case. We then examine the consistency of this ensemble with other lattice and phenomenological findings.

{\em Confinement Phase Transition.---} In this study, we have performed extensive numerical simulations for the statistical ensemble of instanton-dyons as described above. Scanning a wide range of parameter space $(V_c,\zeta_c)$, we  simulate  for each choice the ensemble at different  values of action $S$ (which is basically varying  temperatures).  

\begin{figure}[!t]
\centering
\includegraphics[width=\textwidth]{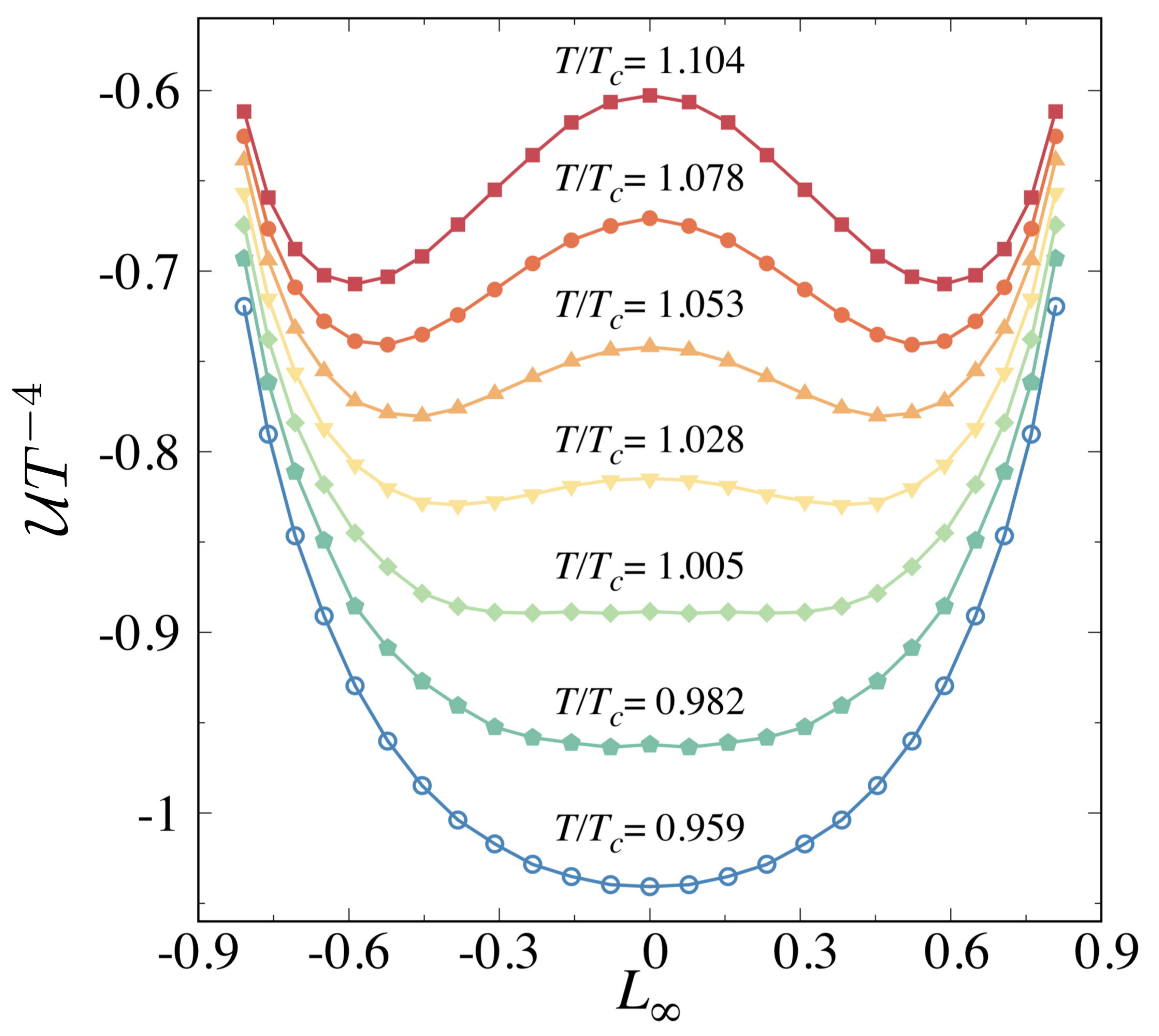}
\caption{The holonomy potential $\mathcal{U}$ as a function of holonomy $L_\infty$ for different action $S$ from larger to smaller values (from top to bottom), or equivalently from higher to lower temperatures. See text for details.  \vspace{-0.15in}} 
\label{fig1}
\end{figure}

A first quantity to examine is the aforementioned holonomy potential at different temperatures. These results for $\mathcal{U}[L_\infty]$ are shown in \cref{fig1} . (For this plot the parameters are chosen as $(V_c=5, \zeta_c=2.4)$, but the observed behavior of the holonomy potential is generically true for other choices of parameters.)  As can be seen, when the action $S$ decreases (i.e. the temperature decreases), the holonomy potential smoothly evolves from a  hump-shape featuring minima away from $L_\infty=0$ toward a valley-shape featuring a minimum at the confining holonomy of $L_\infty=0$. This is characteristic for a  second-order phase transition. In fact, one can identify the critical action $S_c$ (with a corresponding temperature we call $T_c$) where the minimum just moves to $L_\infty=0$. This allows us to do the scaling of temperature via \cref{Eq:2LoopS}. 

Clearly, in the strongly coupled regime (corresponding to smaller $S$ at lower temperature), confinement occurs in the system. Intuitively this result can be understood as follows. With increasingly strong coupling, it costs less  action to create these objects. As a result, the ensemble would eventually become dense enough so that the short-range repulsive force becomes important. In this regime, the holonomy parameter $h$ would prefer to stay at $1/2$ where the $L$- and $M$-sectors are balanced. To see this, imagine that  $h$ would deviate from $1/2$, say $h< 1/2$. In this case the number of $M$-dyons would increase (as their action cost is $hS$) but their size $\sim 1/h$ would also increase thus causing a significant increase of energy cost due to the repulsive interaction. The same argument for $L$-sector applies for the case of $h>1/2$. As a result, when the ensemble becomes dense, the $h=1/2$ point becomes the optimal state of the system.  

\begin{figure}[!t]
\centering
\includegraphics[width=\textwidth]{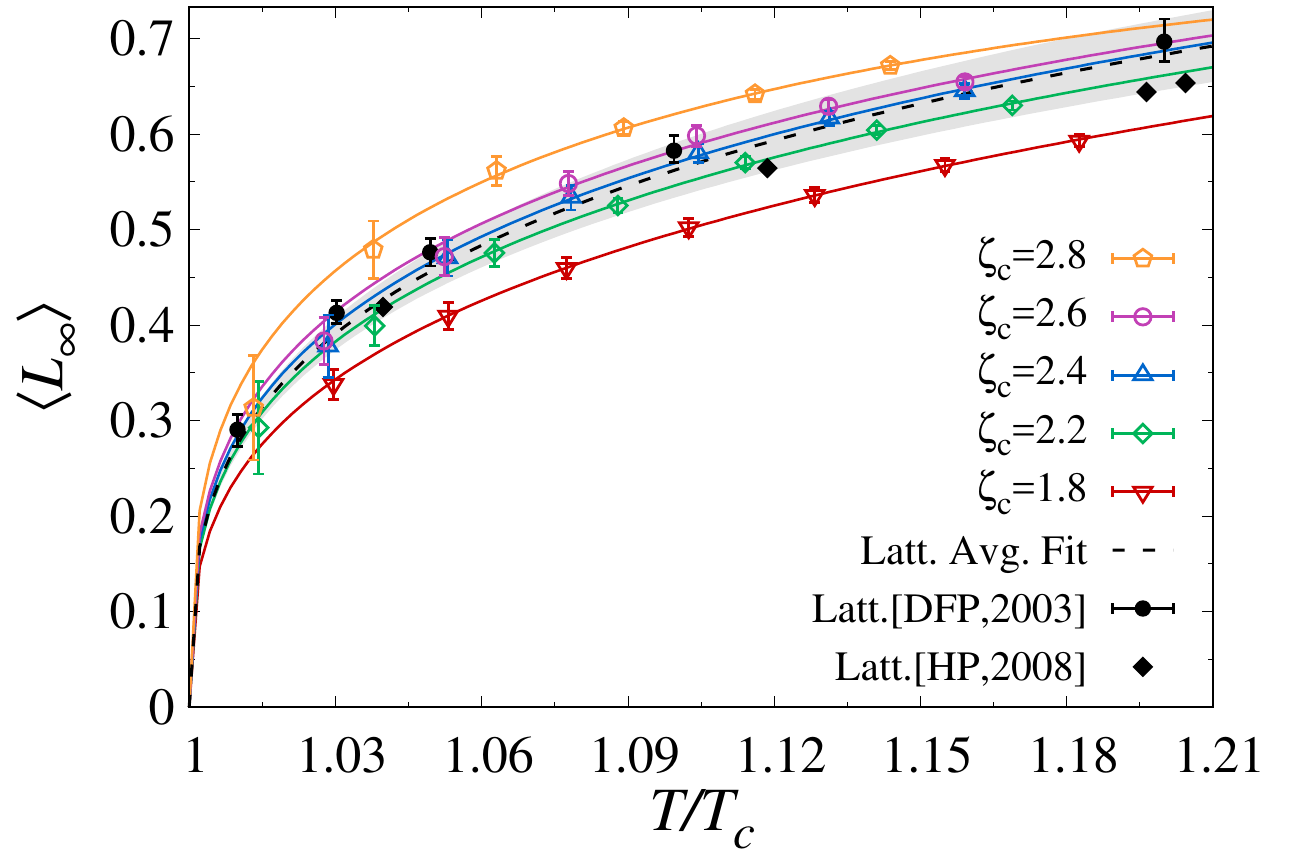}
\caption{The Polyakov loop expectation value $\langle L_\infty \rangle$ versus temperature $T/T_c$. The filled circle and diamond symbols are for lattice results~\cite{Digal:2003jc,Hubner:2008ef}, while the open symbols are from instanton-dyon ensemble calculations with different choices of parameters. See text for details. \vspace{-0.15in}}
\label{fig2}
\end{figure}

With the holonomy potential obtained, one can then determine from its minimum the Polyakov loop expectation value $\langle L_\infty \rangle $ as a function of temperature. As is well known, this is the order parameter for confinement transition in pure Yang-Mills theories. In the $SU(2)$ case, a second-order phase transition is expected with $\langle L_\infty \rangle =0$ at low temperature while non-zero at high temperature. Such dependence $\langle L_\infty \rangle (T)$ for $SU(2)$ Yang-Mills theories has been obtained from lattice simulations, as shown in \cref{fig2} by the filled circle and diamond symbols from two recent lattice works~\cite{Digal:2003jc,Hubner:2008ef}. We use the grey band to indicate the lattice uncertainty as reflected by the minor difference of the two calculations. The results from instanton-dyon ensemble calculations for a few choices of parameters are shown in \cref{fig2} as curves with open symbols. A second-order phase transition is clearly observed in all cases. We scan a wide range of parameter space and compare with lattice results with quantitative $\chi^2$ analysis to constrain the values of $V_c$ and $\zeta_c$. For the repulsive potential strength $V_c$, we find that the results are relatively insensitive to its value  in the range from 5 to 20, with $V_c=5$ giving the best agreement with lattice. The results are however quite sensitive to the range parameter $\zeta_c$, as can be seen from the visible variation of the curves with different $\zeta_c$  in \cref{fig2}.  We see very good agreement with lattice for $\zeta_c \in [2.2,2.6]$ and find the optimal value to be $\zeta_c=2.4$ with  $\chi^2/\text{d.o.f} \approx 1.21$. 

As is well known, it is expected based on the center symmetry of this theory that the Polyakov loop as order parameter would exhibit  a second-order transition with critical scaling behavior near $T_c$ in the same universality class as the 3D Ising model
~\cite{Svetitsky:1982gs,McLerran:1981pb}.  Such behavior is well reproduced by the instanton-dyon ensemble results,  following scaling formula $\langle L_\infty \rangle = b\left(T/Tc-1\right)^{0.3265} \left[1 + d (T/T_c-1)^{0.530}\right]$~\cite{Pelissetto:2000ek,Hubner:2008ef} shown by  the smooth curves in \cref{fig2}. Fitting analysis with mean-field scaling exponent would give a worse fit very close to $T_c$ but work well toward higher temperature.

\begin{figure}[!t]
\centering
\includegraphics[width=\textwidth]{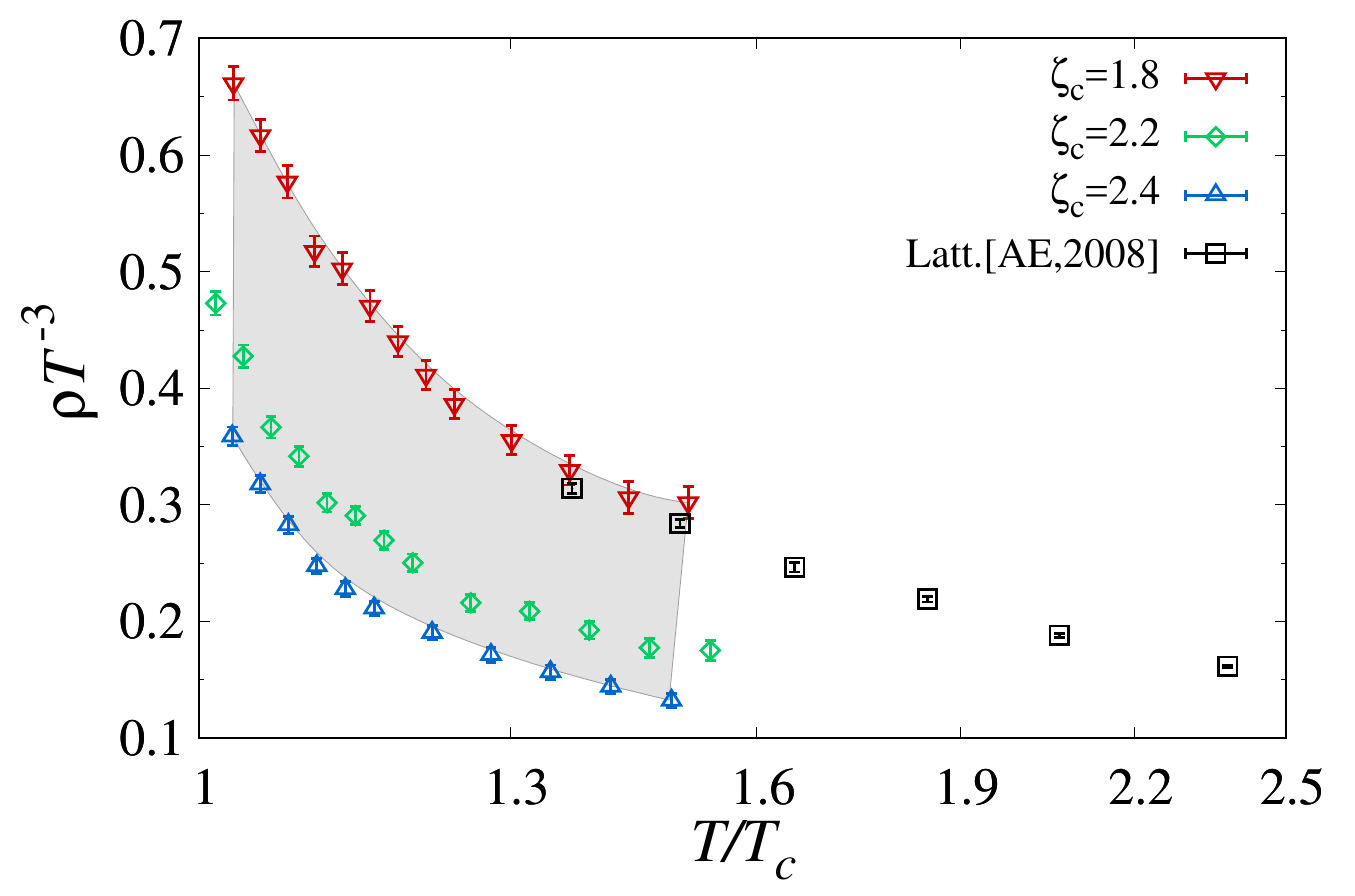}
\caption{The density of chromo-magnetic charges at different temperature from instanton-dyon ensemble, compared with lattice results in \cite{DAlessandro:2007lae}. See text for details. \vspace{-0.15in} } 
\label{fig3}
\end{figure}

{\em Instanton-Dyon Density and Correlations.---} With the key parameters of the instanton-dyon ensemble being characterized above, we now examine its consistency with other relevant information. One such example is  the density of chromo-magnetically  charged objects. This has been studied on the lattice for $SU(2)$ Yang-Mills theory~\cite{DAlessandro:2007lae}. In \cref{fig3}, we compare such density from our instanton-dyon ensemble with that from lattice calculation in \cite{DAlessandro:2007lae}. Results for $\zeta_c$ in the parameter range where the confinement transition can be quantitatively described, are also reasonably consistent the magnetic density from \cite{DAlessandro:2007lae}, with $\zeta_c=1.8$ giving the best agreement.  It may be noted that recent phenomenological studies of jet energy loss observables and heavy flavor transport at the RHIC and  LHC provide interesting evidence for the presence of a chromo-magnetic component in the near-$T_c$ region~\cite{Liao:2006ry,Liao:2008dk,Xu:2014tda,Das:2015ana,Ramamurti:2017zjn,Shi:2018lsf}. The density of magnetic charges  extracted from those studies in the vicinity of $T_c$~\cite{Shi:2018lsf} is about $\rho T^{-3}\simeq (N_c-1) \cdot (0.4\sim 0.6)$, which is also in consistency with   instanton-dyon ensemble results.

Finally, we have also computed the spatial density-density correlations between dyons and anti-dyons in the ensemble. These correlations feature a typical liquid-like behavior in the near-$T_c$ region, with a correlation length on the order of $(0.5\sim 1)\cdot 1/T$. Such observations, again, appear to be viable with experimental observations of the quark-gluon plasma as a nearly perfect liquid at the RHIC and LHC~\cite{Shuryak:2004cy,Gyulassy:2004zy,Muller:2012zq,Shuryak:2014zxa} and with phenomenological studies that suggest the chromo-magnetic component  to play a key role in such observed transport property~\cite{Liao:2006ry,Liao:2008jg,Ratti:2008jz}.   

{\em Conclusion.---} In summary, we have studied a model for describing confinement transition in $SU(2)$ Yang-Mills theory, based on a statistical ensemble of correlated instanton-dyons. This model is shown  to quantitatively describe the lattice data for the temperature dependence of the order parameter. The short-range interaction plays a  crucial role and we have characterized the key parameters of this interaction. The chromo-magnetic charge density as well as the spatial correlations in such ensemble  have also been found to be consistent with known lattice and phenomenological information. We conclude that the correlated instanton-dyon ensemble provides a successful explanation of the confinement mechanism in $SU(2)$ Yang-Mills theory, and may hold the  promise of a similar success for QCD. Interesting and important future tests of this model would include e.g. the Polyakov loop behavior in representations other than the fundamental and the topological susceptibility in the transition region, which will be studied and reported elsewhere.

%
The authors thank R.~Larsen, P.~Petreczky, S.~Shi and in particular E.~Shuryak for helpful discussions and communications. This study is supported in part by NSF (PHY-1913729)  and by the U.S. Department of Energy, Office of Science, Office of Nuclear Physics, within the framework of the Beam Energy Scan Theory (BEST) Topical Collaboration. JL is grateful to the Institute for Advanced Study of Indiana University for partial support. MALR is additionally supported by CONACyT under Doctoral supports Grants No. 669645. YJ is supported by National Natural Science Foundation of China (Grant No. 11875002) and by the Zhuobai Program of Beihang University. This research was supported in part by Lilly Endowment, Inc., through its support for the Indiana University Pervasive Technology Institute, and in part by the Indiana METACyt Initiative. \vspace{-0.15in}
%
%
%

%
\end{document}